\documentclass[11pt,prd,preprintnumbers,nofootinbib,superscriptaddress,notitlepage]{revtex4-1}

\usepackage{graphics}
\usepackage{bm}
\usepackage{cancel}
\usepackage{color}
\usepackage{xcolor}
\usepackage{multirow}
\usepackage{slashed}
\usepackage{array}
\usepackage{amssymb}
\usepackage{graphicx}
\usepackage{mathrsfs}
\usepackage{diagbox}
\usepackage{amsmath}
\usepackage{float}
\usepackage{extarrows}
\usepackage[colorlinks,citecolor=blue,anchorcolor=red,menucolor=red,linkcolor=red,filecolor=red,runcolor=red,urlcolor=red,frenchlinks=red]{hyperref}
\usepackage{subfigure}

\makeatletter
\newcommand{\thickhline}{\noalign {\ifnum 0=`}\fi \hrule height 1pt\futurelet \reserved@a \@xhline}
\newcolumntype{"}{@{\hskip\tabcolsep\vrule width 1pt\hskip\tabcolsep}}
\makeatother

\def\nn{\nonumber}

\def\gev{\rm GeV}

\def\mn{m_N^{}}

\def\pslash{\not{\hbox{\kern-4pt $p$}}}
\def\qslash{\not{\hbox{\kern-4pt $q$}}}
\def\lv{\not{\hbox{\kern-4pt $L$}}}
\def\lsim{\mathrel{\raise.3ex\hbox{$<$\kern-.75em\lower1ex\hbox{$\sim$}}}}
\def\gsim{\mathrel{\raise.3ex\hbox{$>$\kern-.75em\lower1ex\hbox{$\sim$}}}}
\def\ifmath#1{\relax\ifmmode #1\else $#1$\fi}

\def\slash{\!\!\!/}

\begin{document}

\title{Measuring CP violation in rare $W$ decays at the LHC}

\author{Peng-Cheng Lu}
\email{pclu@sdu.edu.cn}
\affiliation{School of Physics, Shandong University, Jinan, Shandong 250100, China}

\author{Zong-Guo Si}
\email{zgsi@sdu.edu.cn}
\affiliation{School of Physics, Shandong University, Jinan, Shandong 250100, China}

\author{Zhe Wang}
\email{wzhe@mail.sdu.edu.cn}
\affiliation{School of Physics, Shandong University, Jinan, Shandong 250100, China}

\author{Xing-Hua Yang}
\email{yangxinghua@sdut.edu.cn}
\affiliation{School of Physics and Optoelectronic Engineering, Shandong University of Technology, Zibo, Shandong 255000, China}

\begin{abstract}
Heavy Majorana neutrinos beyond the standard model can simultaneously explain the origin of tiny neutrino masses and matter-antimatter asymmetry in our Universe.
The existence of heavy Majorana neutrinos will also lead to lepton number violation and the rare lepton-number-violating $W$ decays are possible.
With contributions from two different Majorana neutrinos, a nonzero CP asymmetry may be generated from the rate difference between $W$ decay and its CP-conjugate process.
The aim of this paper is to investigate the prospects for measuring CP violation in rare $W$ decays via Majorana neutrinos at the LHC.
Our calculations show that the induced CP asymmetry is independent of the Majorana neutrino mass for $10~\gev < \mn < 70~\gev$.
Such a CP asymmetry if observed, would in turn provide unambiguous evidence of new physics beyond the standard model.

\end{abstract}


\maketitle

\section{INTRODUCTION}\label{sec1}

The standard model (SM) of particle physics has proved to be extremely successful in describing all the known fundamental particles and interactions. However, there are still several questions that remain unexplained.
For example, one of the main mysteries is the origin of tiny neutrino masses. Within the framework of SM, neutrinos are predicted to be massless. However, the discovery of neutrino oscillations has firmly indicated that neutrinos are massive particles and lepton flavors are mixed~\cite{ParticleDataGroup:2020ssz}.
Another important mystery is how to explain the observed matter-antimatter asymmetry of the Universe. It is argued that the baryon asymmetry in our Universe can be characterized by the baryon-to-photon density ratio $\eta=n_B/n_\gamma$.
From a careful analysis of recent Planck measurements of the cosmic microwave background, the value of $\eta$ has been determined to a good degree of accuracy: $\eta \simeq (6.12 \pm 0.03) \times 10^{-10}$~\cite{Planck:2018vyg}, which is just too large compared to the SM expectation.
To go beyond the SM, the simplest way to accommodate neutrino masses is to introduce $n$ right-handed Majorana neutrino fields $N_{\rm R}$, which can couple to the left-handed neutrino fields through Yukawa interactions to form Dirac mass terms $\overline{\nu_{\rm L}}M_{\rm D}N_{\rm R}$.
As being SM gauge singlets, the introduced right-handed neutrino fields $N_{\rm R}$ are also allowed to couple to their charge conjugate fields to constitute Majorana mass terms $\overline{N^c_{\rm L}}M_{\rm R}N_{\rm R}$.
This is known as the famous type-I seesaw mechanism and the tiny neutrino masses can be given by $M_\nu \approx -M_{\rm D} M^{-1}_{\rm R} M^{\rm T}_{\rm D}$~\cite{Minkowski:1977sc,
Mohapatra:1979ia}.
In this canonical seesaw mechanism, the smallness of the left-handed neutrinos can be attributed to the heaviness of the right-handed Majorana neutrinos.
As for the origin of the baryon asymmetry in our Universe, one popular explanation is leptogenesis mechanism~\cite{Fukugita:1986hr}. In this case, the CP-violating and out-of-equilibrium decays of the heavy Majorana neutrinos thermally produced in the early Universe may first generate lepton number asymmetry, and the latter may be subsequently converted into baryon number asymmetry through sphaleron processes~\cite{Manton:1983nd,Klinkhamer:1984di}.

The key point behind the seesaw mechanism and the associated leptogenesis mechanism is the existence of the heavy Majorana neutrinos, which will cause the violation of lepton number by two units ($\Delta L =2$) simultaneously and the lepton-number-violating processes such as neutrinoless double-beta decay ($0\nu\beta\beta$)~\cite{Furry:1939qr,Elliott:2004hr} may be possible.The underlying process with $\Delta L =2$ can be generically expressed as a $W$ decay via Majorana neutrino exchange, which can be identified by the signature of same-sign dilepton in the final state.
This rare $W$ decay, to be specific, $W^- \rightarrow \ell_{1}^{-} N \rightarrow \ell_{1}^{-}\ell_{2}^{-}(q\bar{q}^\prime)^{+}$, has been well studied in the literature (for a review, see e.g. Refs.~\cite{Deppisch:2015qwa,Antusch:2016ejd,Cai:2017mow}).
For example, in the mass range of $M_N < M_W$, the $\Delta L =2$ same-sign dilepton production signal has been explored in rare meson decays~\cite{
Atre:2005eb,Cvetic:2010rw,
Wang:2014lda,Dong:2013raa,
Cvetic:2016fbv}, tau lepton decays~\cite{Gribanov:2001vv,Kobach:2014hea,Yuan:2017xdp} and even top quark decays~\cite{Bar-Shalom:2006osy,Si:2008jd,Delepine:2012nea,Liu:2019qfa}.
For heavy Majorana neutrinos with masses above $M_W$, the same signal has been extensively investigated at various collider experiments, such as the electron-positron colliders~\cite{Buchmuller:1991tu,Gluza:1993gf,
Cvetic:1998vg,delAguila:2005ssc}, electron-proton colliders~\cite{Buchmuller:1990vh,Ingelman:1993ve,Liang:2010gm,Lindner:2016lxq,Li:2018wut,Gu:2022muc} and proton-proton
colliders\cite{Keung:1983uu,Datta:1993nm,Han:2006ip,Atre:2009rg,Chao:2009ef,Fuks:2020att}.
The difference between the rates of $W^- \rightarrow \ell_{1}^{-}\ell_{2}^{-}(q\bar{q}^\prime)^{+}$ and its CP-conjugate process $W^+ \rightarrow \ell_{1}^{+}\ell_{2}^{+}(\bar{q}q^\prime)^{-}$ may induce a nonzero CP asymmetry, which arises from the significant interference of different Majorana neutrinos.
The generated CP violation effects can serve as smoking gun for new physics beyond the SM.
A great deal of work has been done for trying to measure the CP violation effects in the decays of meson~\cite{
Cvetic:2015naa,
Cvetic:2020lyh,Godbole:2020jqw,Zhang:2020hwj} and tau lepton~\cite{Zamora-Saa:2016ito,Tapia:2019coy}.
Recently, Ref.~\cite{Najafi:2020dkp} explored the CP violation in rare $W$ decays at the LHC, but the CP violation effect produced in this case is influenced by the initial parton distribution functions in proton.
To evade this conflict, in a recent work~\cite{Lu:2021vzj}, we investigated the possibility for measuring CP violation in $t\bar{t}$ pair production and rare decay at the LHC. As the $W^-$ and $W^+$ bosons respectively arise from the decays of $\bar{t}$ and $t$ quarks, they have the same quantity.
Different from the previous studies, in this paper, we explore the discovery prospects for measuring CP violation in $W^+W^-$ pair production and rare decay at the LHC, where the $W^-$ and $W^+$ bosons are produced directly from $pp$ collisions.
In principle, the number $n$ of introduced right-handed Majorana neutrinos in seesaw mechanism is a free parameter. Since two neutrino mass-squared differences between light neutrinos have been observed, $n \geq 2$ is needed.
A nonzero direct CP asymmetry also requires the existence of at least two different Majorana neutrinos.
For illustration, we take two heavy Majorana neutrinos in consideration, and the general case with more Majorana neutrinos can be analyzed in a similar way.

This paper is organized as follows. A theoretical framework for heavy Majorana neutrinos is briefly introduced in Section~\ref{sec2}. The rare $W$ decays via Majorana neutrinos are discussed in Section~\ref{sec3}. In Section~\ref{sec4}, we explore the experimental prospects for measuring CP vioaltion at the LHC. Finally, we conclude in Section~\ref{sec5}.

\section{Heavy Majorana neutrinos beyond SM }\label{sec2}

Throughout this paper, we only consider a minimal extension of the SM by introducing two right-handed Majorana neutrinos.
In the notation of Ref.~\cite{Atre:2009rg}, the flavor eigenstates $\nu_{\ell}$ (with $\ell= e, \mu, \tau$) of three active neutrinos can be expressed by the mass eigenstates of light and heavy Majorana neutrinos:
\begin{equation}
\label{1}
 \nu_{\ell L} = \sum_{m=1}^{3} V_{\ell m} \nu_{m L} + \sum_{m'=1}^{2} R_{\ell m^\prime} N^{c}_{m' L} \; .
\end{equation}
Therefore, the weak charged-current interaction Lagrangian can now be written in terms of the mass eigenstates as follows:
\begin{align}
\label{2}
-\mathcal{L}_{\rm cc}
= \frac{g}{\sqrt{2}} W^{+}_{\mu} \sum_{\ell=e}^{\tau} \sum_{m=1}^{3} V_{\ell m}^{\ast} \overline{\nu_{m}} \gamma^{\mu} P_{L} \ell
+ \frac{g}{\sqrt{2}} W^{+}_{\mu} \sum_{\ell=e}^{\tau} \sum_{m'=1}^{2} R_{\ell m^\prime}^{\ast} \overline{N^{c}_{m^\prime}} \gamma^{\mu} P_{L} \ell + \text{h.c.} \; .
\end{align}
Here, $V_{\ell m}$ is just the Pontecorvo-Maki-Nakagawa-Sakata (PMNS) matrix~\cite{Pontecorvo:1957cp,Maki:1962mu} responsible for neutrino oscillations.
Note that $R_{\ell m^\prime}$ describes the light-heavy neutrino mixing and can be generally parameterized as
\begin{equation}
\label{3}
R_{\ell m^\prime}=\left|R_{\ell m^\prime}\right|e^{i\phi_{\ell m^\prime}} , ~~~ \ell= e, \mu, \tau, ~~~ m^\prime=1, 2 \; .
\end{equation}
It is worth pointing out that the heavy Majorana neutrinos in the conventional type-I seesaw mechanism are usually too heavy (e.g. at order of GUT-scale) and their mixings with light neutrinos are severely suppressed.
However, there also exists some low-scale seesaw scenarios with heavy Majorana neutrino masses being much lower and the strength of light-heavy neutrino mixings being large enough (for a review, see e.g. Refs.~\cite{Xing:2009in,Chen:2011de}).
In this paper, we will adopt a model-independent phenomenological approach by taking the heavy neutrino masses $m_{N}$ and light-heavy neutrino mixings $R_{\ell m^\prime}$ as free parameters.

Constraints on the free parameters $m_{N}$ and $R_{\ell m^\prime}$ mentioned above can be derived from experimental observations, and a detailed summary can be found in Ref.~\cite{Deppisch:2015qwa}.
In our calculations, to be conservative, we take
\begin{eqnarray}
\label{4}
\left|R_{e i}\right|^2 = 1.0 \times 10^{-7}  \; ,~~ \left|R_{\mu i}\right|^2 = \left|R_{\tau i}\right|^2 = 1.0 \times 10^{-5} \; , ~~~  i = 1, 2 \; ,
\end{eqnarray}
for $10~\gev < \mn < 70~\gev$, which are consistent with $0\nu\beta\beta$-decay searches~\cite{GERDA:2020xhi}, a global fit to lepton flavor and electro-weak precision data~\cite{Fernandez-Martinez:2016lgt}, a reanalysis of the Large Electron Positron (LEP) collider data~\cite{DELPHI:1996qcc} as well as direct searches by the Large Hadron Collider (LHC) experiments~\cite{CMS:2018jxx}.

\section{CP violation in rare $W$ decays}\label{sec3}

\begin{figure}[!htbp]
\begin{center}
\subfigure[]{\label{fig1a}
\includegraphics[width=0.4\textwidth]{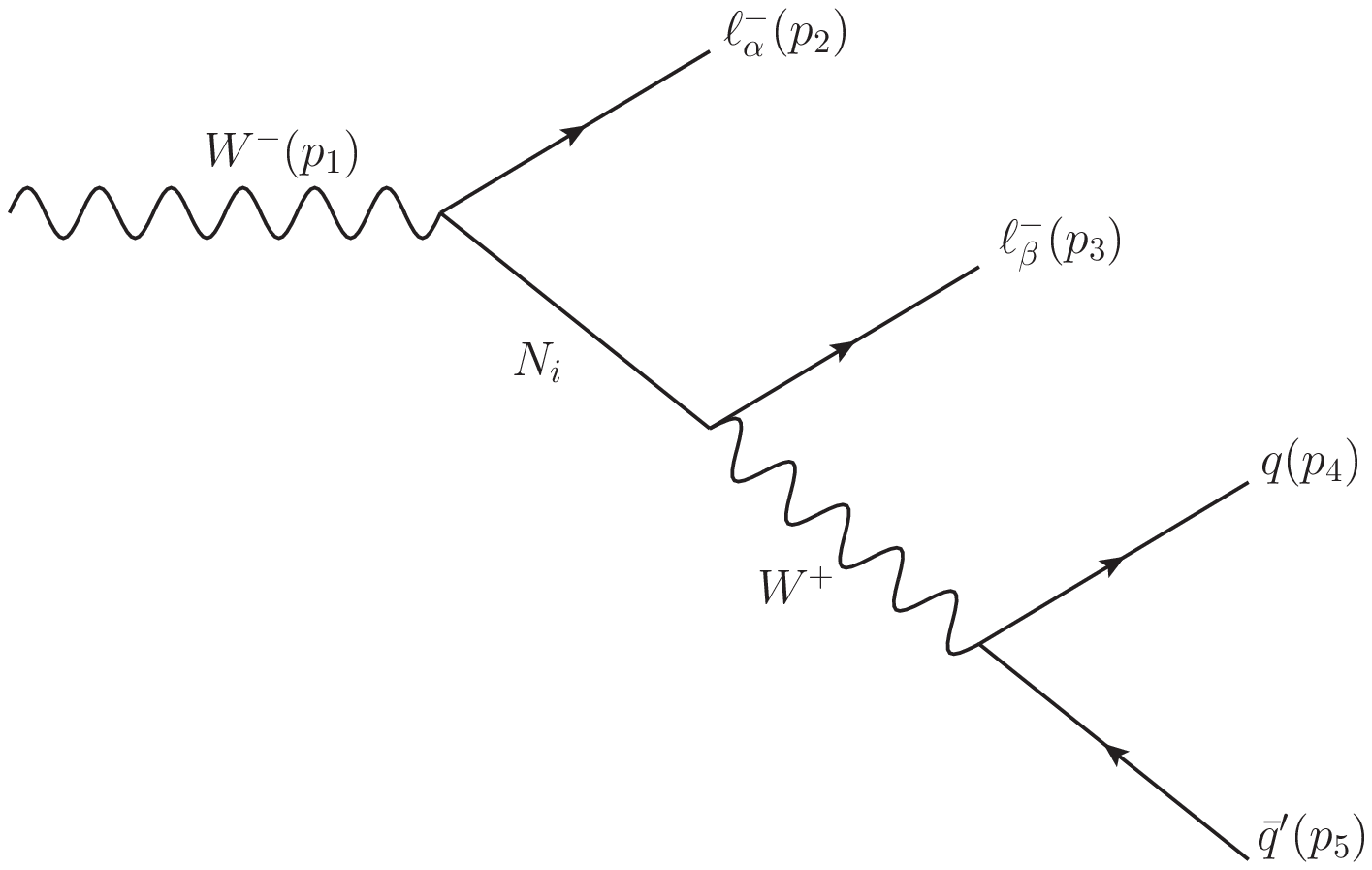} }
\hspace{-0.5cm}~
\subfigure[]{\label{fig1b}
\includegraphics[width=0.4\textwidth]{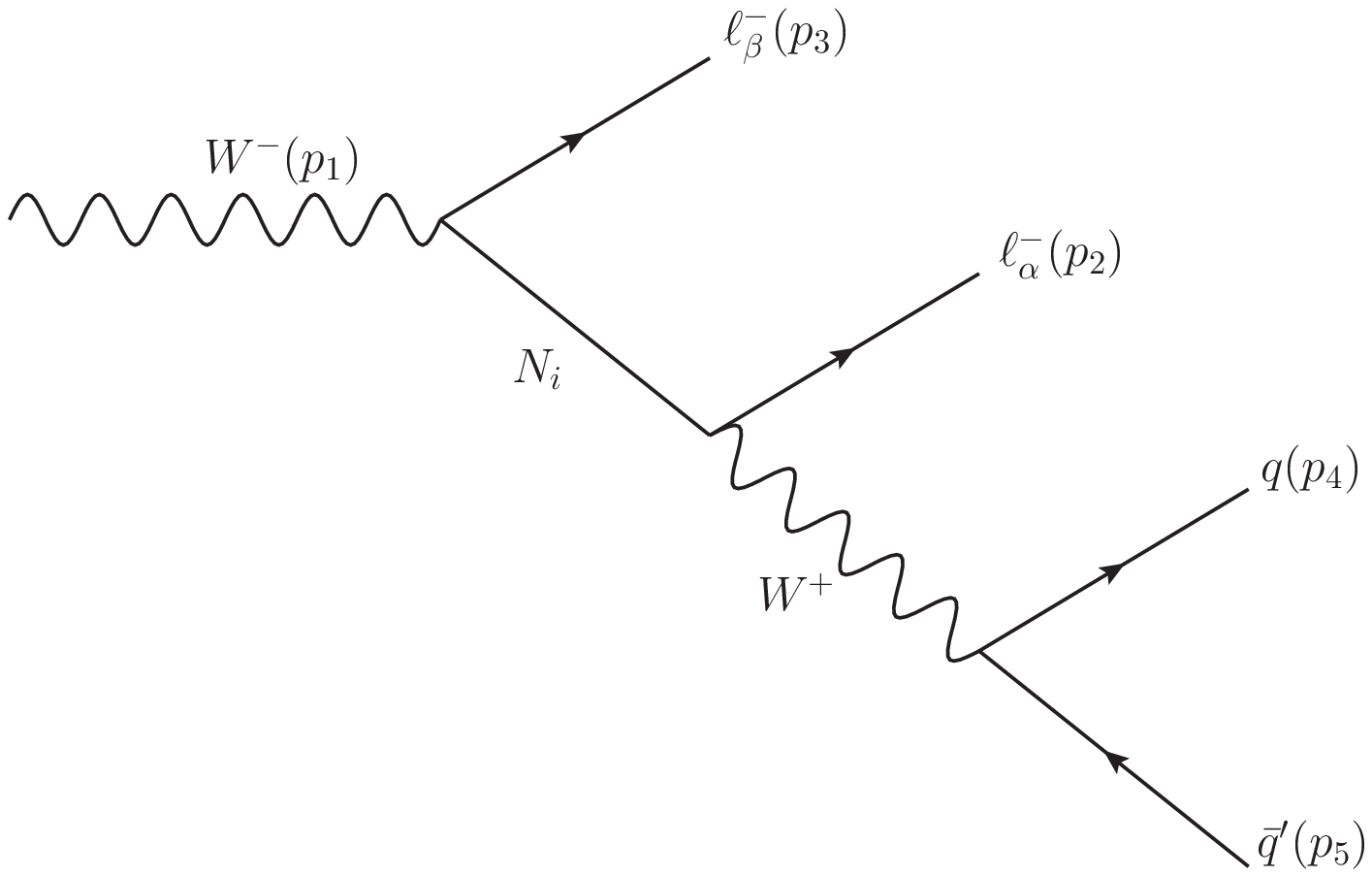} }
\caption{Feynman diagrams for rare $W^-$ decay via heavy Majorana neutrino exchange.}\label{fig1}
\end{center}
\end{figure}

The rare $W$ decays can be induced by heavy Majorana neutrinos.
Here, we consider both the rare lepton-number-violating decay of $W^-$ and $W^+$ bosons via two intermediate on-shell Majorana neutrinos $N_{i}$ (with $i=1, 2$) (depicted in Fig.~\ref{fig1}):
\begin{align}
\label{5}
W^-(p_1) &\to \ell_{\alpha}^{-}(p_2) + N_{i}(p_N) \to \ell_{\alpha}^{-}(p_2) + \ell_{\beta}^{-}(p_3) + q(p_4) + \bar{q}^\prime(p_5) \; , \nn \\
W^+(p_1) &\to \ell_{\alpha}^{+}(p_2) + N_{i}(p_N) \to \ell_{\alpha}^{+}(p_2) + \ell_{\beta}^{+}(p_3) + \bar{q}(p_4) + q^\prime(p_5) \; ,
\end{align}
where $\alpha, \beta = e, \mu, \tau$. Note that $p_1$, $p_2$, etc. denote the four-momentum of the corresponding particles, respectively.
The squared matrix elements averaged (summed) over the initial (final) particles for the process in Eq.~(\ref{5}) can be obtained as follows:
\begin{align}
\label{6}
\overline{\left|{\cal M}_{\ell_\alpha^\pm\ell_\beta^\pm}\right|^2} =& \frac{g^6}{ m_W^2}\left|V_{qq^\prime}\right|^2\left(1-\frac{1}{2}\delta_{\alpha\beta}\right)
\left|D_{W}\left(p_w^2\right)\right|^2 \nn \\
& \times \left\{m_{N_1}^{2}\left|R_{\alpha 1}R_{\beta 1}\right|^2{\cal T}_1 + m_{N_2}^{2}\left|R_{\alpha 2}R_{\beta 2}\right|^2{\cal T}_2 \right. \nn \\
& \left. + m_{N_1} m_{N_2} \left|R_{\alpha 1}R_{\alpha 2}R_{\beta 1}R_{\beta 2}\right|{\rm Re}\left[e^{\pm i\Delta\phi}{\cal T}_{12}\right] \right\} \; ,
\end{align}
where $p_w=p_4+p_5$. $D_W\left(p^2\right) =1/( p^2-m_W^2+im_W\Gamma_W)$ where $m_W$, $\Gamma_W$ are the mass and total decay width of $W$ boson.
 Note that $V_{qq^\prime}$ is the Cabibbo-Kobayashi-Maskawa (CKM) matrix element~\cite{Cabibbo:1963yz,Kobayashi:1973fv} and set to be diagonal with unit entries for simplicity.
Also note that $R_{\alpha i }$ (with $\alpha = e, \mu, \tau$ and $i = 1, 2$) is the light-heavy neutrino mixing matrix element defined in  Eq.~(\ref{3}) and the complex phase $\Delta\phi=\phi_{\alpha 2} - \phi_{\alpha 1} +  \phi_{\beta 2} -  \phi_{\beta 1}$ coming from the significant interference between $N_1$ and $N_2$ can serve as a new source of CP violation. The explicit expressions of ${\cal T}_i$ ($i = 1, 2$) and ${\cal T}_{12}$ are shown in Appendix~\ref{appA}.

Furthermore, the corresponding differential decay width can then be expressed as
\begin{eqnarray}
\label{7}
d\Gamma_{W^\pm \to \ell_{\alpha}^\pm\ell_{\beta}^\pm (q \bar{q}^\prime)^\mp}=\frac{1}{2m_W}\overline{\left|{\cal M_{\ell_\alpha^\pm\ell_\beta^\pm}}\right|^2} d{\cal L}ips_4 \; ,
\end{eqnarray}
with $d{\cal L}ips_4$ being the Lorentz invariant phase space of the four final particles.
The decay modes of heavy Majorana neutrino have been well studied in Ref.~\cite{Atre:2009rg}. As the decay width of heavy Majorana neutrino is much smaller than its mass in the mass range of our interest, the narrow-width approximation (NWA)~\cite{Uhlemann:2008pm} can then be applied.
Therefore, the total decay width for the process in Eq.~(\ref{5}) can be factorized as follows:
\begin{eqnarray}
\label{8}
\Gamma_{W^\pm \to \ell_{\alpha}^\pm\ell_{\beta}^\pm (q \bar{q}^\prime)^\mp}
\approx \Gamma_{W^\pm \to \ell_{\alpha}^\pm N_i}\cdot{\rm Br}\left(N_i \to \ell_{\beta}^\pm (q \bar{q}^\prime)^\mp\right)
= \left(2-\delta_{\alpha\beta}\right)\cdot S_{\alpha\beta}\cdot\Gamma_{0} \; ,
\end{eqnarray}
where $\Gamma_{0}$ is a function of Majorana neutrino mass and the ``effective mixing parameter" $S_{\alpha\beta}$ is defined as
\begin{eqnarray}
\label{9}
S_{\alpha\beta}=\frac{\left|R_{\alpha i}R_{\beta i}\right| ^{2}}{\sum_{\ell=e}^{\tau} \left|R_{\ell i}\right| ^{2}} \; .
\end{eqnarray}

\begin{figure}[!htbp]
\begin{center}
\subfigure[]{\label{fig2a}
\includegraphics[width=0.4\textwidth]{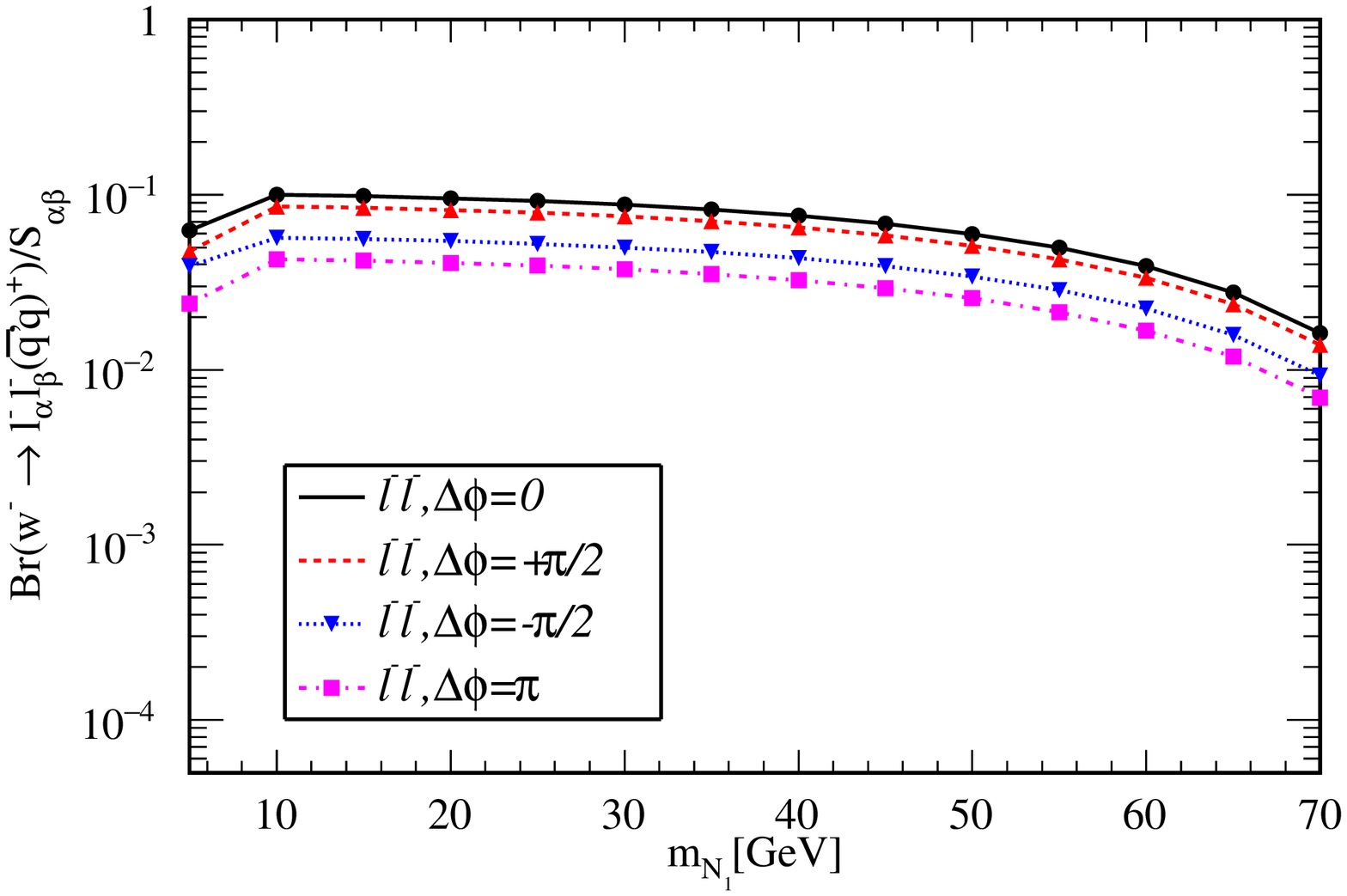} }
\hspace{-0.5cm}~
\subfigure[]{\label{fig2b}
\includegraphics[width=0.4\textwidth]{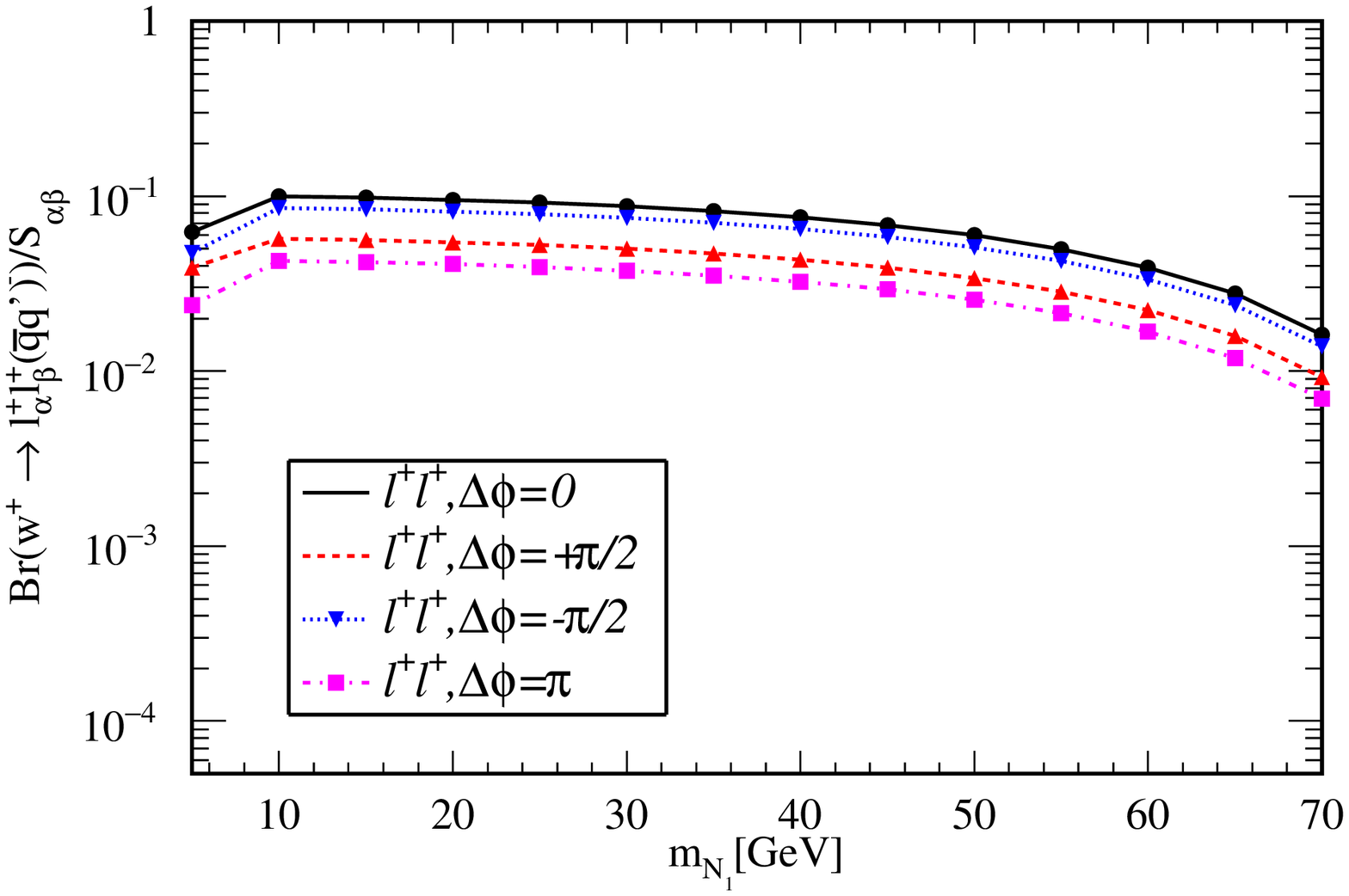} }
\caption{The normalized branching ratio of (a) $W^- \to \ell_{\alpha}^- \ell_{\beta}^- (q \bar{q}^\prime)^+$ and (b) $W^+ \to \ell_{\alpha}^+ \ell_{\beta}^+ (\bar{q}q^\prime)^{-}$ versus Majorana neutrino mass $m_{N_{1}}$ for $\Delta\phi=0, +\pi/2, -\pi/2, \pi$.}\label{fig2}
\end{center}
\end{figure}

\begin{figure}[!htbp]
\begin{center}
\subfigure[]{\label{fig3a}
\includegraphics[width=0.4\textwidth]{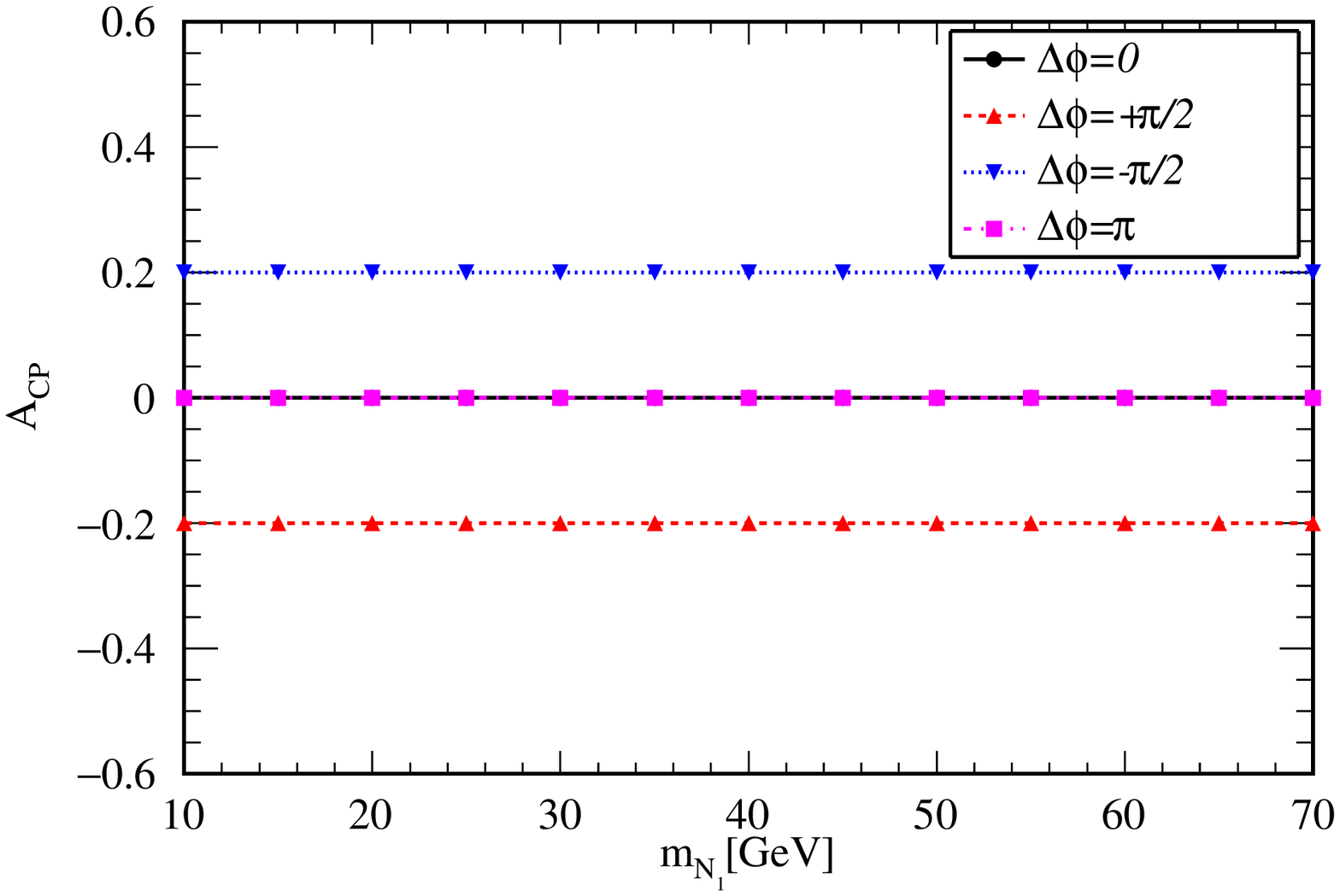} }
\hspace{-0.5cm}~
\subfigure[]{\label{fig3b}
\includegraphics[width=0.4\textwidth]{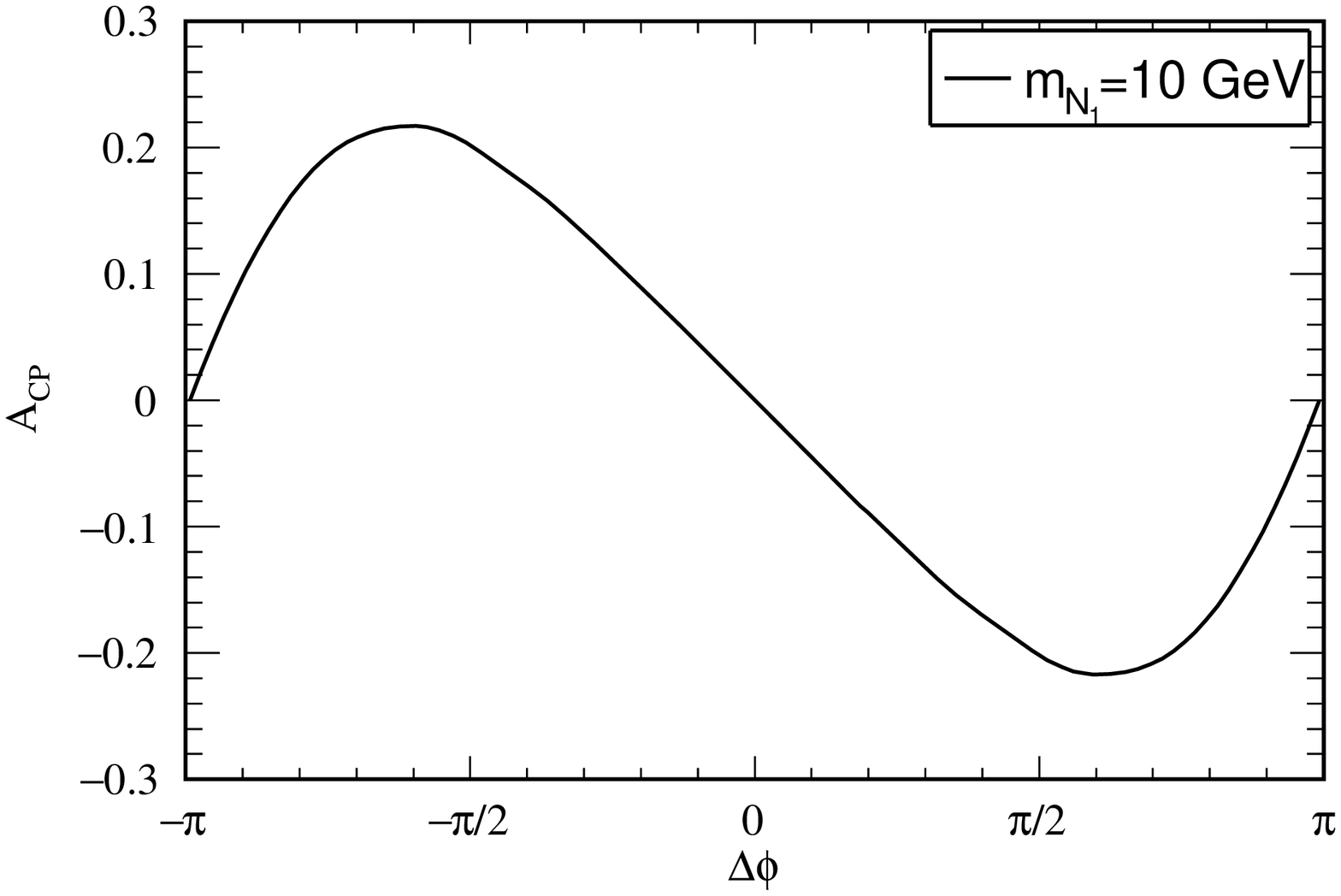} }
\caption{(a) The value of ${\cal A}_{\rm CP}$ as a function of $m_{N_{1}}$ for $\Delta\phi=0, +\pi/2, -\pi/2, \pi$. (b) The value of ${\cal A}_{\rm CP}$ as a function of $\Delta\phi$ for $m_{N_{1}} = 10~\gev$. }\label{fig3}
\end{center}
\end{figure}

In our numerical calculations, we employ the following approximations:
\begin{eqnarray}
\label{10}
m_{N_{2}} = m_{N_{1}} + \Gamma_{N_{1}}/2  \; , ~~ \Gamma_{N_{2}} \approx \Gamma_{N_{1}} \; .
\end{eqnarray}
The normalized branching ratio of $W^- \to \ell_{\alpha}^- \ell_{\beta}^- (q \bar{q}^\prime)^+$ and $W^+ \to \ell_{\alpha}^+ \ell_{\beta}^+ (\bar{q}q^\prime)^{-}$ as a function of $m_{N_{1}}$ are shown in Fig.~\ref{fig2}.
For illustration, the CP phase difference is set to $\Delta\phi=0, +\pi/2, -\pi/2, \pi$.
It is found that, the normalized branching ratio decreases slowly as the Majorana neutrino mass increases  for $10~\gev < m_{N_{1}} < 70~\gev$. When the Majorana neutrino mass is close to the $W$ boson mass, the normalized branching ratio falls off sharply.
The difference between the rates of $W^- \rightarrow \ell_{1}^{-}\ell_{2}^{-}(q\bar{q}^\prime)^{+}$ and $W^+ \rightarrow \ell_{1}^{+}\ell_{2}^{+}(\bar{q}q^\prime)^{-}$ may induce the CP asymmetry, which can be defined as
\begin{eqnarray}
\label{11}
{\cal A}_{\rm CP} = \frac{\Gamma_{W^- \to \ell_{\alpha}^- \ell_{\beta}^- (q \bar{q}^\prime)^+} - \Gamma_{W^+ \to \ell_{\alpha}^+ \ell_{\beta}^+ (\bar{q}q^\prime)^{-}}}
{\Gamma_{W^- \to \ell_{\alpha}^- \ell_{\beta}^- (q \bar{q}^\prime)^+} + \Gamma_{W^+ \to \ell_{\alpha}^+ \ell_{\beta}^+ (\bar{q}q^\prime)^{-}}} \; .
\end{eqnarray}
As also can be seen in Eq.~(\ref{6}), this CP asymmetry arises from the interference of contributions from two different heavy Majorana neutrinos.
Therefore, in order to generate such a CP asymmetry, the following two necessary conditions must be satisfied: (i) the existence of at least two heavy Majorana neutrinos and (ii) the non-zero CP phase difference $\Delta\phi$.
The numerical results of ${\cal A}_{\rm CP}$ versus the Majorana neutrino mass $m_{N_{1}}$ for various values of $\Delta\phi$ are shown in Fig.~\ref{fig3a}.
Taking $m_{N_{1}} = 10~\gev$ for example, the value of ${\cal A}_{\rm CP}$ as a function of $\Delta\phi$ is also displayed in Fig.~\ref{fig3b}.
It is obvious that, for fixed $\Delta\phi$, ${\cal A}_{\rm CP}$ is independent of the Majorana neutrino mass in the mass range of $10~\gev < \mn < 70~\gev$.
Furthermore, it can also be found that the CP asymmetry vanishes for $\Delta\phi=0, \pi$, and ${\cal A}_{\rm CP} \to -{\cal A}_{\rm CP}$ for $\Delta\phi \to -\Delta\phi$.
When $\Delta\phi\approx\pm3\pi/5$, the maximal value of $|{\cal A}_{\rm CP}|_{\rm max}\approx0.22$ can be reached.

\section{CP violation in $W^+W^-$ pair production and rare decay at the LHC}\label{sec4}

\begin{figure}[!htbp]
\begin{center}
\subfigure[]{\label{fig4a}
\includegraphics[width=0.4\textwidth]{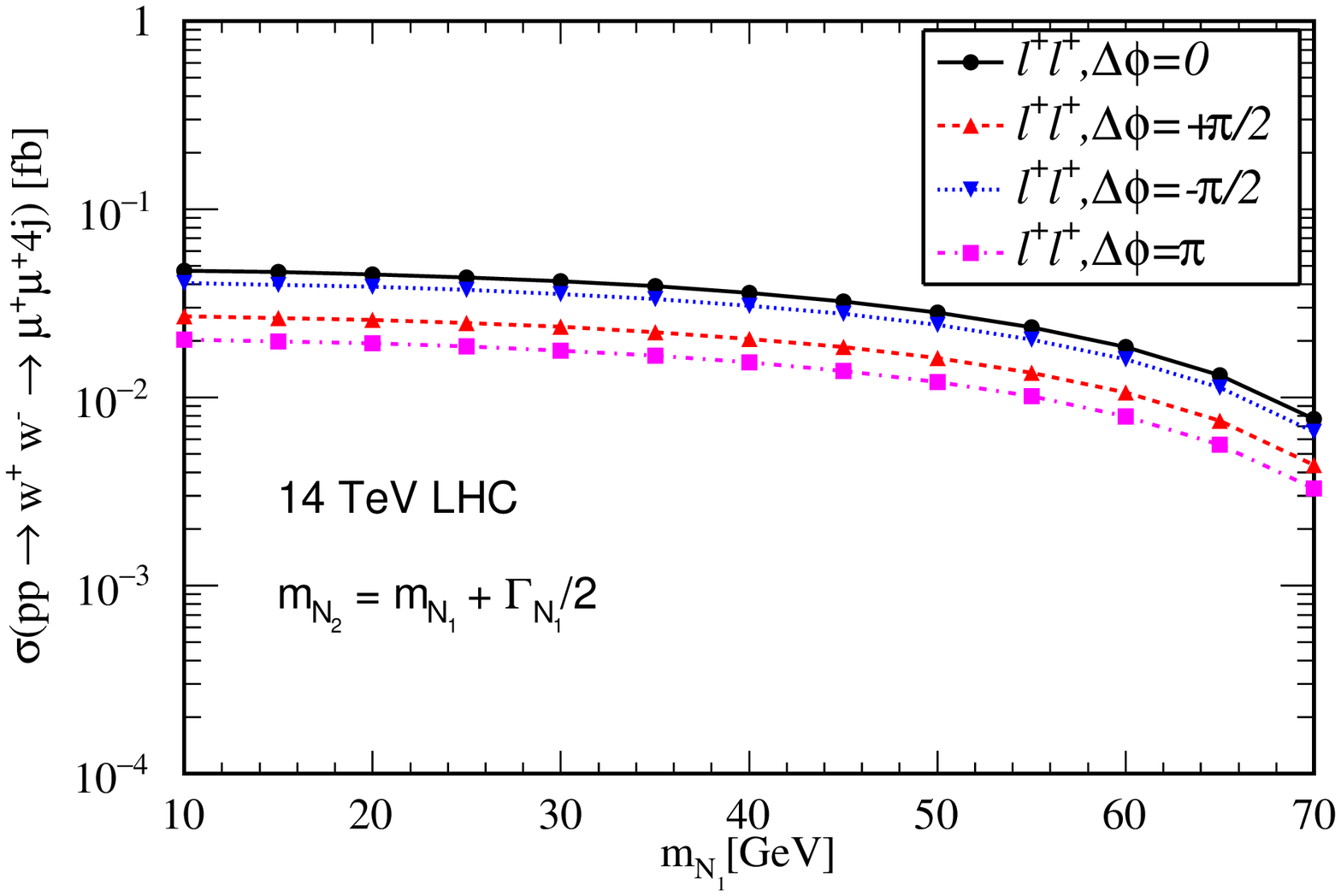} }
\hspace{-0.5cm}~
\subfigure[]{\label{fig4b}
\includegraphics[width=0.4\textwidth]{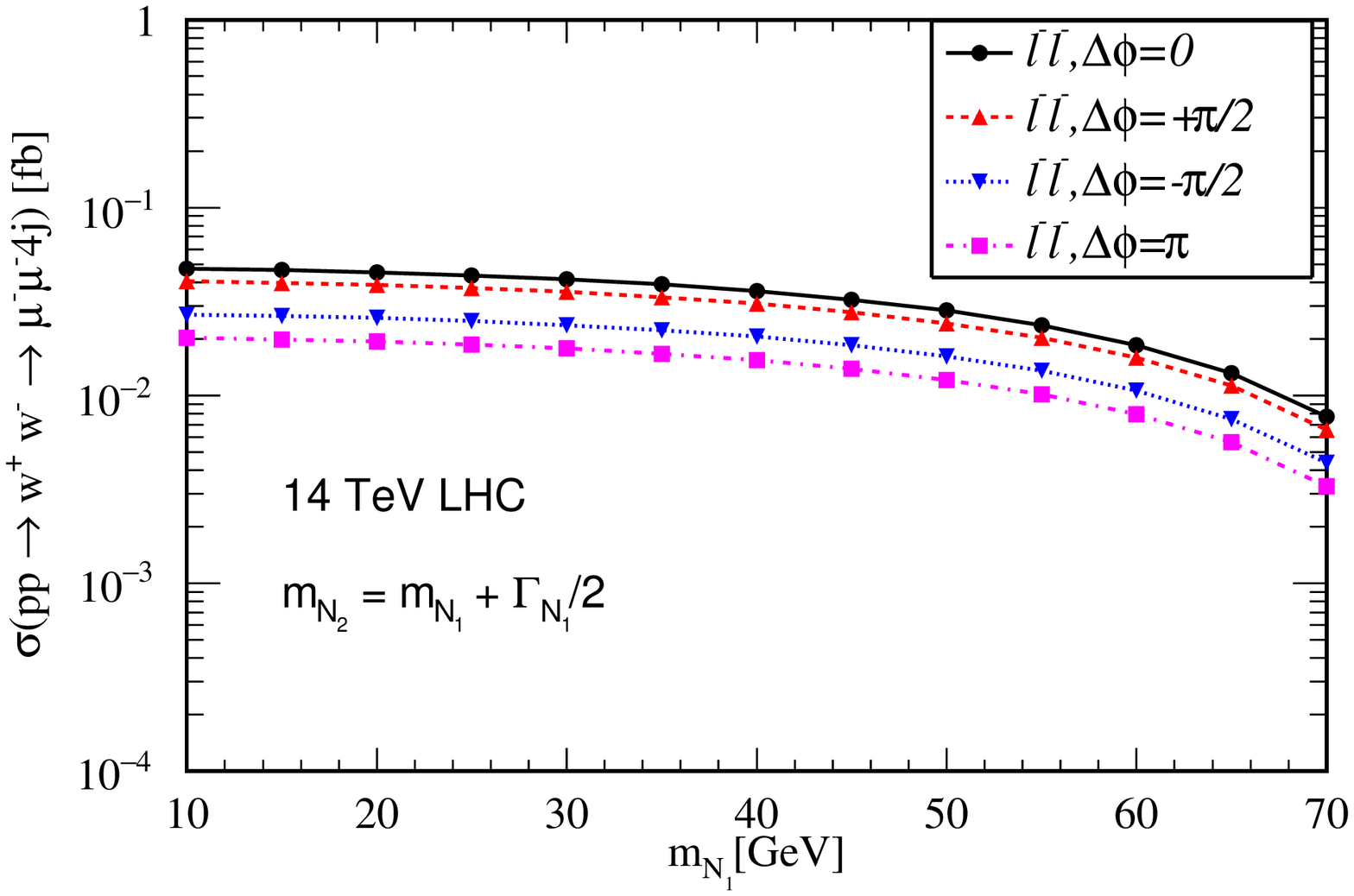} }
\hspace{-0.5cm}~
\subfigure[]{\label{fig4c}
\includegraphics[width=0.4\textwidth]{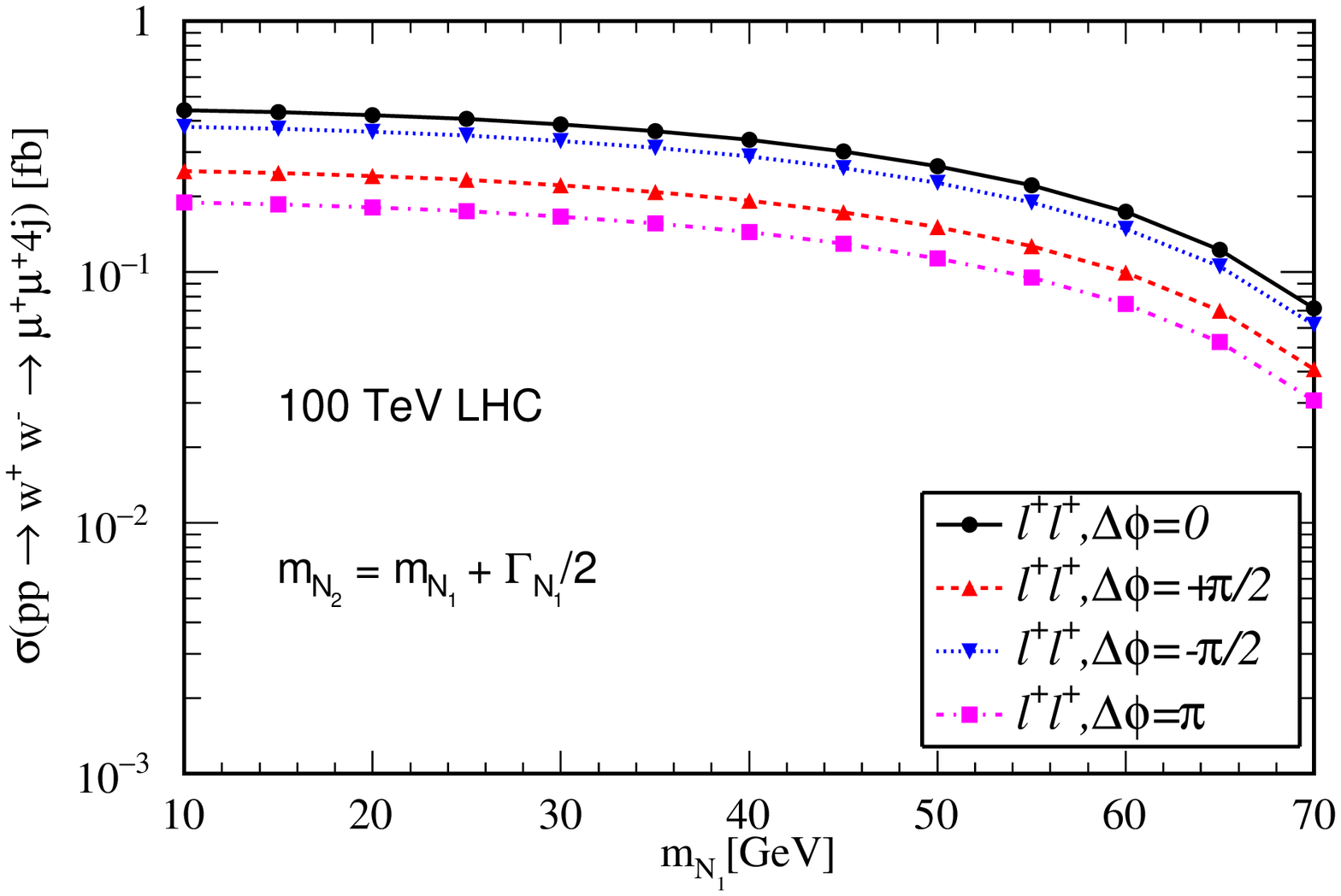} }
\hspace{-0.5cm}~
\subfigure[]{\label{fig4d}
\includegraphics[width=0.4\textwidth]{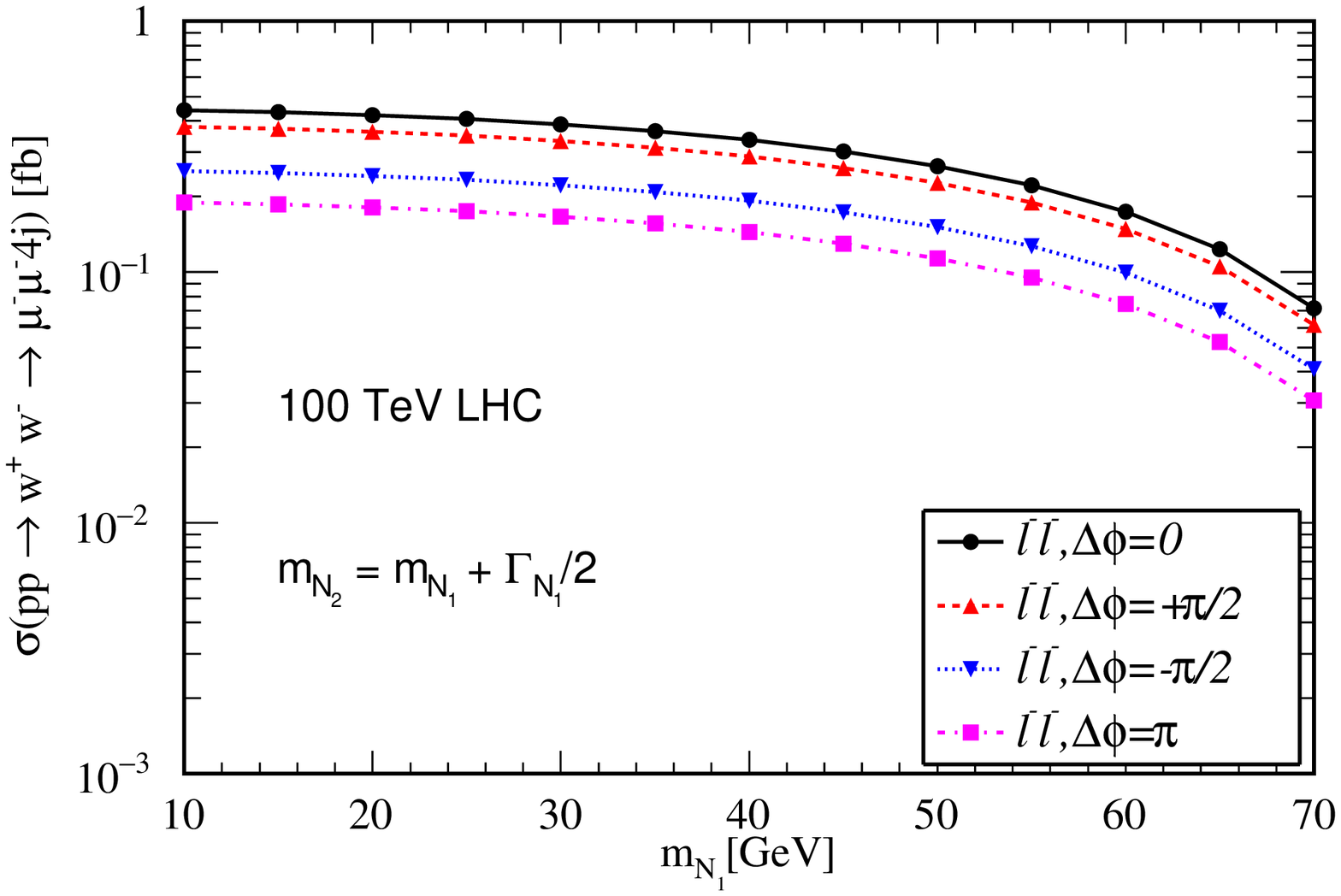} }
\caption{The total cross sections for (a) $pp \to W^+W^- \to \mu^+ \mu^+ 4j$ and (b) $pp \to W^+W^- \to \mu^- \mu^- 4j$ at 14 TeV LHC, (c) $pp \to W^+W^- \to \mu^+ \mu^+ 4j$ and (d) $pp \to W^+W^- \to \mu^- \mu^- 4j$ at 100 TeV LHC versus Majorana neutrino mass $m_{N_{1}}$ with $\Delta\phi=0, +\pi/2, -\pi/2, \pi$. }\label{fig4}
\end{center}
\end{figure}

With unprecedented high energy and high luminosity, the LHC will offer a great opportunity to probe new physics beyond the SM.
In this paper, we explore the prospects for measuring CP violation in rare $W$ decays at the LHC and consider the following process
\begin{eqnarray}
\label{12}
pp \to W^\pm W^\mp \to \ell^\pm \ell^\pm  + 4j \; ,
\end{eqnarray}
where the $W^\pm W^\mp$ pairs are produced directly from $pp$ collisions.
In the following numerical calculations, we restrict our study to the same-sign dimuon production channel and employ CTEQ6L1~\cite{Pumplin:2002vw} for the parton distribution functions in proton.
In Fig.~\ref{fig4}, the total cross sections for the process in Eq.~(\ref{12}) are shown as a function of $m_{N_{1}}$ for $\Delta\phi=0, +\pi/2, -\pi/2, \pi$ at 14 TeV and 100 TeV LHC, respectively.
With the integrated luminosity of $\cal{L}$=300~${\rm fb}^{-1}$, there only a few events produced for $m_{N_{1}} < 70~\gev$ at 14 TeV LHC, and more than ten events can be produced at 100 TeV LHC.

\begin{figure}[!htbp]
\begin{center}
\subfigure[]{\label{fig5a}
\includegraphics[width=0.4\textwidth]{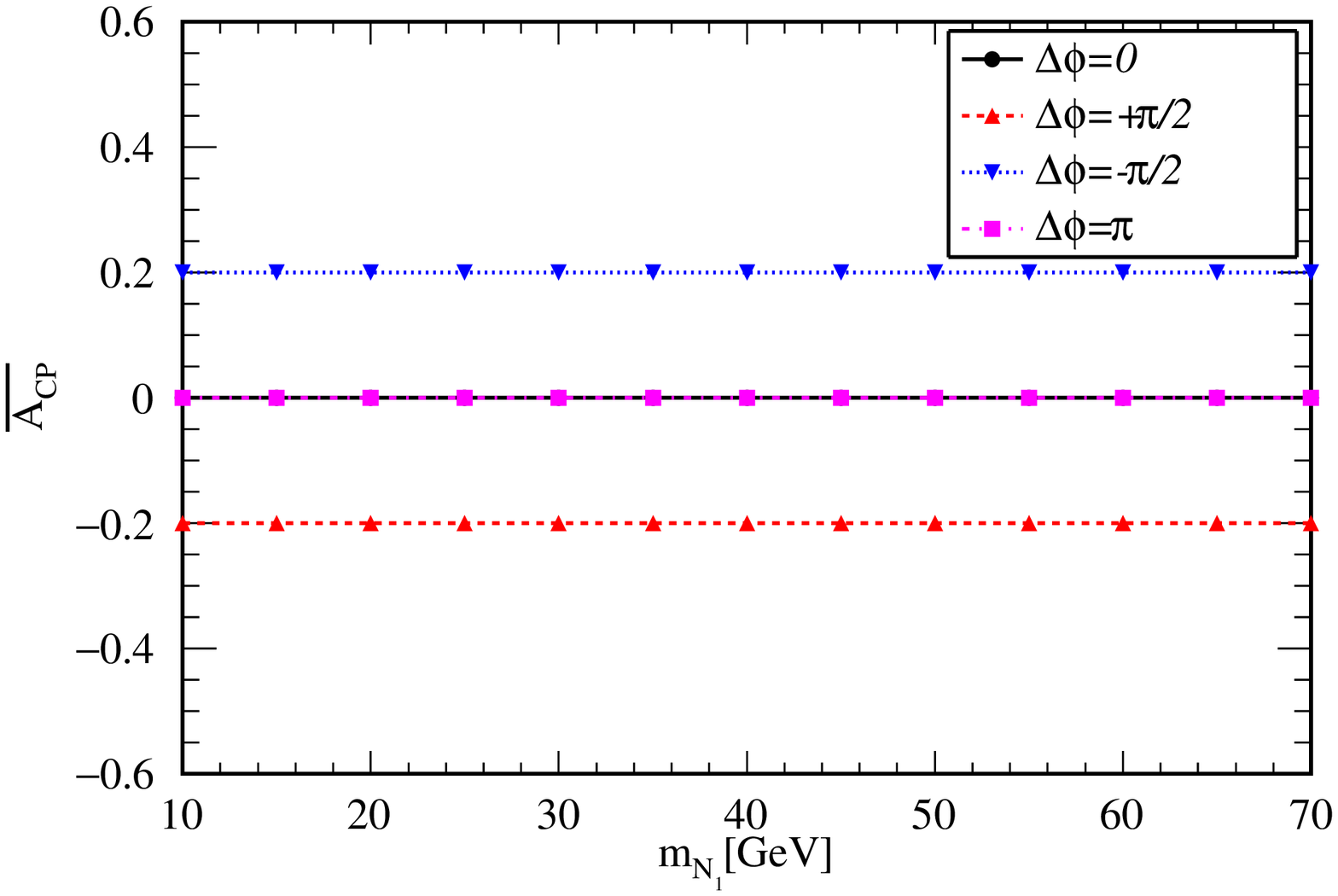} }
\hspace{-0.5cm}~
\subfigure[]{\label{fig5b}
\includegraphics[width=0.4\textwidth]{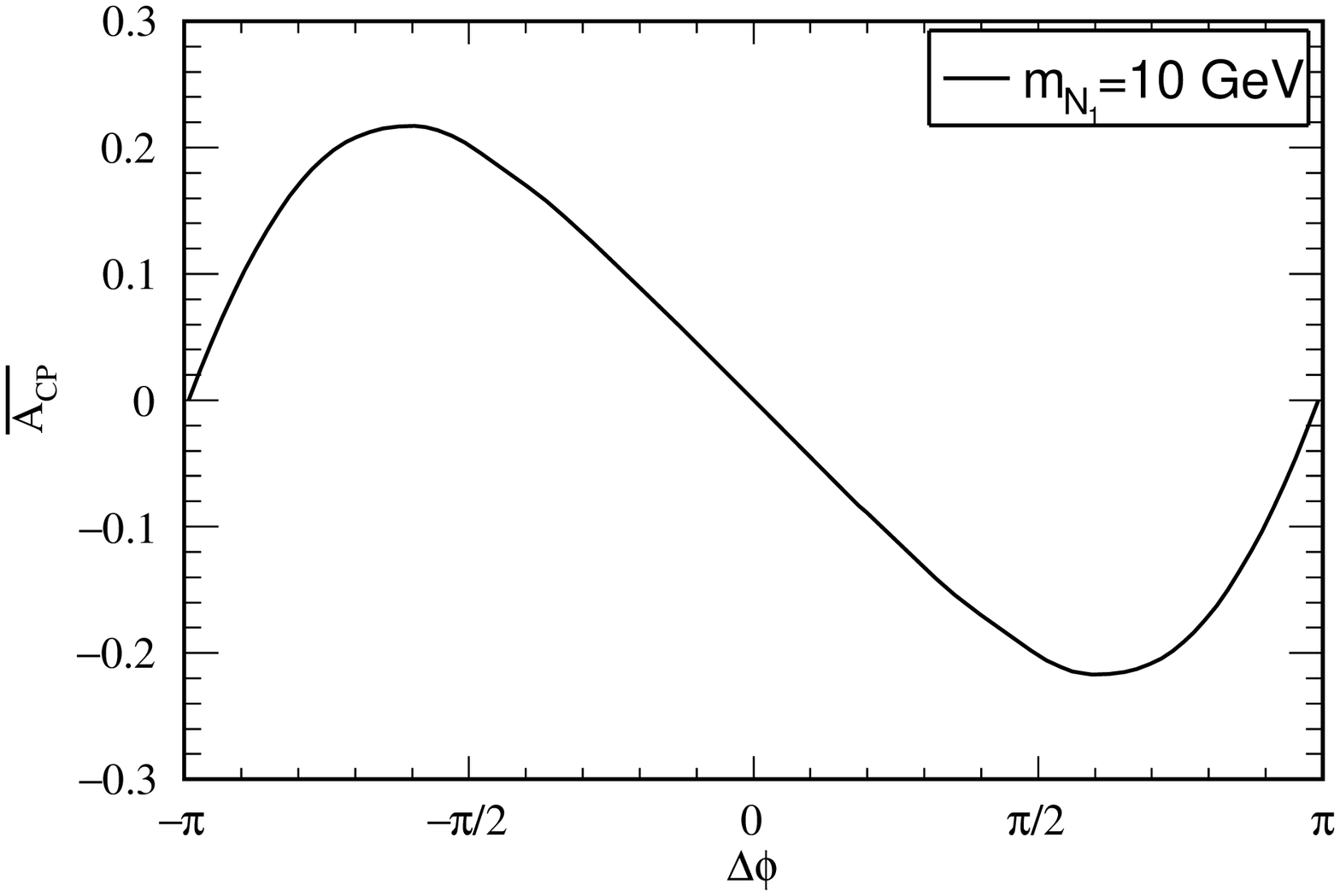} }
\caption{(a) The value of $\overline{{\cal A}_{\rm CP}}$ as a function of $m_{N_{1}}$ for $\Delta\phi=0, +\pi/2, -\pi/2, \pi$. (b) The value of $\overline{{\cal A}_{\rm CP}}$ as a function of $\Delta\phi$ for $m_{N_{1}} = 10~\gev$. }\label{fig5}
\end{center}
\end{figure}

Analogously, the difference between $\sigma(pp \to \mu^+ \mu^+ 4j)$ and $\sigma(pp \to \mu^- \mu^- 4j)$ can also lead to a non-zero CP asymmetry, which can be expressed as
\begin{eqnarray}
\label{13}
\overline{{\cal A}_{\rm CP}} = \frac{\sigma(pp \to \mu^+ \mu^+ 4j) - \sigma(pp \to \mu^- \mu^- 4j)}
{\sigma(pp \to \mu^+ \mu^+ 4j) + \sigma(pp \to \mu^- \mu^- 4j)} .
\end{eqnarray}
In this case, the underlying CP violation effect is similarly caused by the rate difference between rare $W^-$ decay and its CP-conjugate process.
Therefore, the CP asymmetry defined in Eq.~(\ref{13}) is equivalent to that given in Eq.~(\ref{11}).
In Fig.~\ref{fig5}, we display the value of $\overline{{\cal A}_{\rm CP}}$ with respect to $m_{N_{1}}$ and $\Delta\phi$, respectively.
As expected, the CP asymmetry in Fig.~\ref{fig5} behaves almost the same as that in Fig.~\ref{fig3}.

In order to simulate detector response, the lepton and jet energies are smeared according to the assumption of Gaussian resolution parametrization
\begin{eqnarray}
\label{14}
\frac{\delta(E)}{E} = \frac{a}{\sqrt{E}}\oplus b,
\end{eqnarray}
where $\delta(E)/E$ represents the energy resolution and $\oplus$ denotes a sum in quadrature.
In our calculations, we take $a=5\%$, $b=0.55\%$ for leptons and $a=100\%$, $b=5\%$ for jets, respectively~\cite{CMS:2007sch,ATLAS:2009zsq}.
The isolated leptons and jets are identified by angular separation, which can be defined as
\begin{eqnarray}
\label{15}
\Delta R_{ij} = \sqrt{\Delta \phi^2_{ij}+\Delta\eta^2_{ij}} \; ,
\end{eqnarray}
with $\Delta\phi_{ij}$ ($\Delta \eta_{ij}$) being the azimuthal angle (rapidity) difference of the corresponding particles.

As the jets from rare $W^\pm$ decay $W^\pm \to \ell_{}^\pm \ell_{}^\pm jj$ are much softer than those in the hadronic decay of $W^\mp \to  j j$, the former two jets can be merged into one fat jet.
Therefore, we only require three jets ($n_j=3$) in the final state of our signal process.
To quantify the signal observability, we impose the following basic acceptance cuts on leptons and jets (referred as cut-I)
\begin{eqnarray}
\label{16}
p_{T}^{\ell} > 10~{\rm GeV} \; , \; |\eta^{\ell}| < 2.8 \; , \; p_{T}^{j} > 15~{\rm GeV} \; , \; |\eta^{j}| < 3.0 \; , \; 0.4 < \Delta R_{\ell j} < 3.5 \; , \;n_j=3  \; .
\end{eqnarray}
Our signal process in Eq.~(\ref{12}) consists of two same-sign dileptons and three jets.
To purify the signal, the missing transverse energy are required to satisfy (referred as cut-II)
\begin{eqnarray}
\label{17}
 {E\slash}_{T} < 20~{\rm GeV} \; .
\end{eqnarray}

For our signal process, the main backgrounds in SM come from $pp \to W^{\pm}W^{\pm}W^{\mp}W^{\mp}$, $pp \to W^{\pm}W^{\pm}W^{\mp}j$ and $pp \to W^{\pm}W^{\pm}W^{\mp}Z$.
To be specific, the SM backgrounds are simulated by \textsf{MadGraph5\_aMC@NLO}~\cite{Alwall:2014hca}.
The parton shower is performed with \textsf{Pythia-8.2}~\cite{Sjostrand:2006za}.
Jets-clustering is done
with the anti-$k_t$ algorithm~\cite{Cacciari:2008gp}.

Comparing our signal process with the backgrounds, we further fully reconstruct the two Ws. One W boson which decay hadronically can be reconstructed from the two jets ($j_1,j_2$). The invariant mass of this two jets is closest to $m_W$. After reconstructing one W boson, the remaining ingredients are grouped to reconstruct the other W boson. We adopt the following cut (referred as cut-III)
\begin{eqnarray}
\label{17-1}
 |M_{j_1j_2}-m_W| < 20~{\rm GeV} \; , \; |M_{\ell \ell j_3}-m_W| < 20~{\rm GeV}\; ,
\end{eqnarray}
where $j_3$ refers to the left jet and $\ell \ell$ are the two same-sign dileptons.

\begin{figure}[!htbp]
\begin{center}
\subfigure[]{\label{fig6a}
\includegraphics[width=0.4\textwidth]{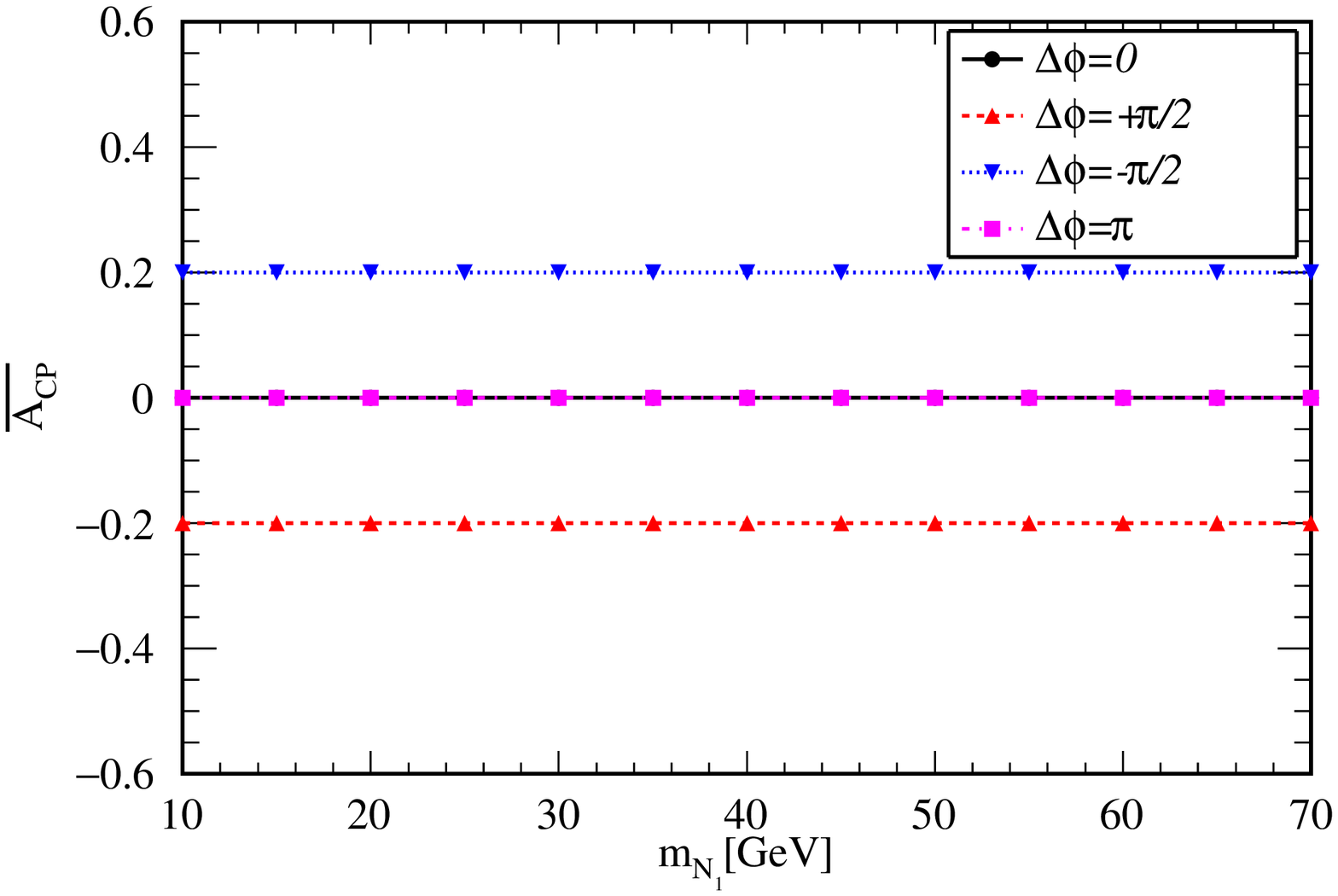} }
\hspace{-0.5cm}~
\subfigure[]{\label{fig6b}
\includegraphics[width=0.4\textwidth]{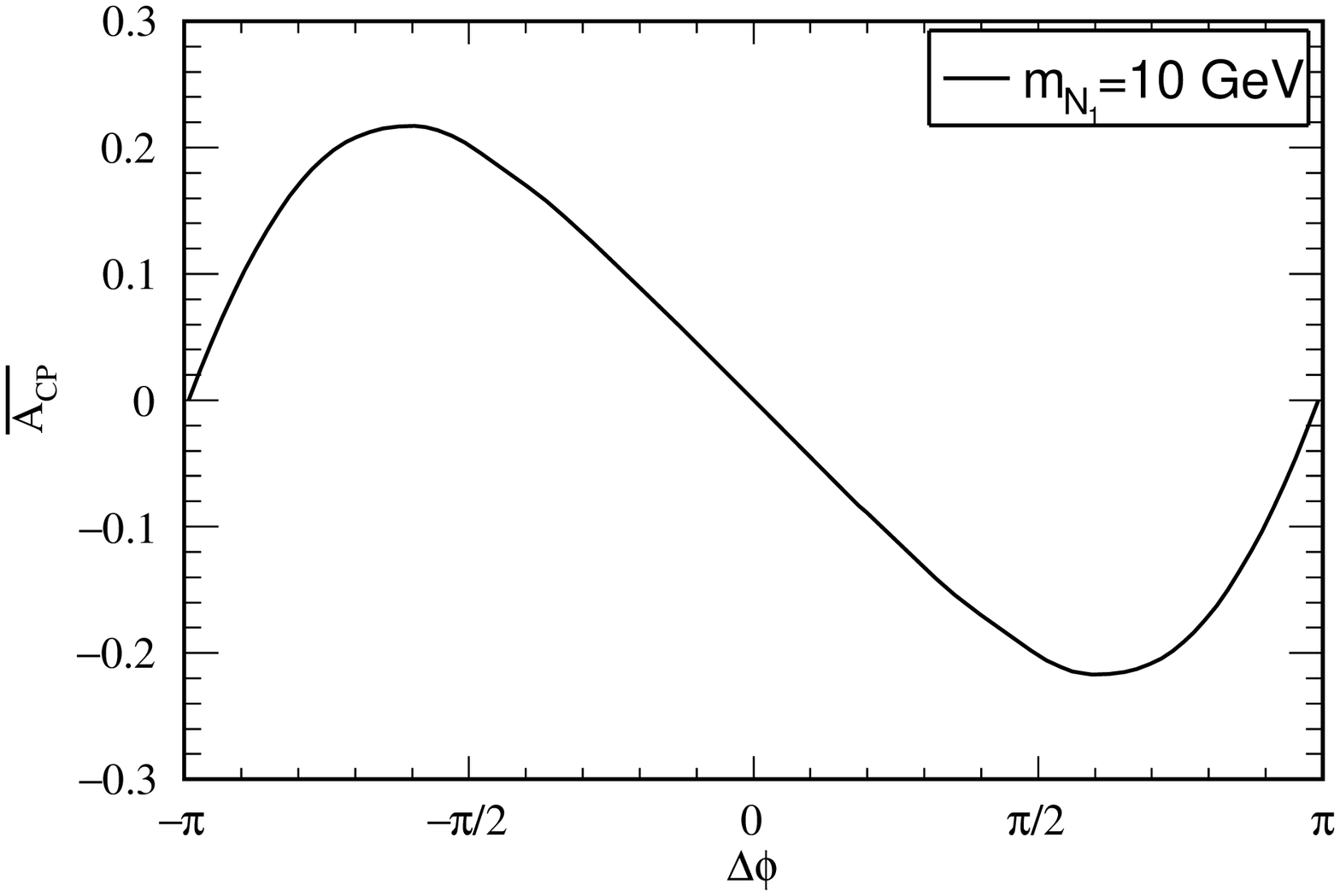} }
\caption{(a) The value of $\overline{{\cal A}_{\rm CP}}$ after all cuts as a function of $m_{N_{1}}$ for $\Delta\phi=0, +\pi/2, -\pi/2, \pi$. (b) The value of $\overline{{\cal A}_{\rm CP}}$ after all cuts as a function of $\Delta\phi$ for $m_{N_{1}} = 10~\gev$. }\label{fig6}
\end{center}
\end{figure}

After implementing all the above cuts, we obtain the total cross sections for the signal and background processes at 14 TeV and 100 TeV, respectively. For illustration, we have used $m_{N_{1}}=10~\gev$ and $\Delta\phi=\pi/2$.
At 14 TeV, the signal cross section after all cuts is only $2.32\times 10^{-3}$~$\mathrm{fb}^{}$, which is too small to be detected experimentally.
We list the total cross sections for the signal and background processes at 100 TeV LHC in Table~\ref{table1}. The statistical significance $S/\sqrt{B}$ with integrated luminosity of $\cal{L}$=300~$\mathrm{fb}^{-1}$ and $\cal{L}$=3000~$\mathrm{fb}^{-1}$ are also given, where $S$ and $B$ denote respectively the signal and background event numbers after all cuts.
 It is shown that the signal cross section after all cuts is remains $1.40\times 10^{-2}$~$\mathrm{fb}^{}$ at 100 TeV LHC and the corresponding statistical significance can reach 4.34 (13.74) with $\cal{L}$=300~$\mathrm{fb}^{-1}$ ($\cal{L}$=3000~$\mathrm{fb}^{-1}$), which will offer us a great opportunity to explore CP violation effects in rare $W$ decays at the LHC.
After adopting all the kinematic cuts, we display respectively $\overline{{\cal A}_{\rm CP}}$ versus $m_{N_{1}}$ and $\Delta\phi$ at 100 TeV LHC in Fig.~\ref{fig6}.
It can be found that the CP asymmetry is almost unchanged after all the selection cuts.

\begin{table}[!htbp]
  \caption{The cross sections for the signal and background processes at 100 TeV LHC after all cuts. Also shown is the statistical significance $S/\sqrt{B}$ with integrated luminosity of $\cal{L}$=300~$\mathrm{fb}^{-1}$ and $\cal{L}$=3000~$\mathrm{fb}^{-1}$. For illustration, we have used $m_{N_{1}}=10~\gev$ and $\Delta\phi=\pi/2$. }\label{table1}
  \centering
  \begin{tabular}{|l|c|c|c|c|c|}
  \hline
       & \multicolumn{4}{c|}{100 TeV} \\
  \cline{2-5}
   & $\sigma_S$ [$\mathrm{fb}$]  & $\sigma_{pp\to W^\pm W^\pm W^\mp W^\mp}$ [$\mathrm{fb}$]
   & $\sigma_{pp\to W^\pm W^\pm W^\mp j}$ [$\mathrm{fb}$]& $\sigma_{pp\to W^\pm W^\pm W^\mp Z}$ [$\mathrm{fb}$]\\
  \hline
  Cut-I &    $2.29\times 10^{-2}$  & $2.9\times 10^{-2}$ & $5.56$ & $1.53\times 10^{-2}$\\
  \hline
  Cut-II &   $1.98\times 10^{-2}$  & $5.45\times 10^{-4}$ &$ 1.59\times 10^{-1}$ &$2.88\times 10^{-4}$  \\
  \hline
  Cut-III &    $1.40\times 10^{-2}$  & $1.21\times 10^{-5}$ &$3.10\times 10^{-3}$  &$3.02\times 10^{-6}$ \\
  \hline
    $S/\sqrt{B}$ with $\cal{L}$=300~$\mathrm{fb}^{-1}$    &  \multicolumn{4}{c|}{4.34} \\
  \hline
  $S/\sqrt{B}$ with $\cal{L}$=3000~$\mathrm{fb}^{-1}$  &  \multicolumn{4}{c|}{13.74} \\
  \hline
\end{tabular}
\end{table}

\section{Summary}\label{sec5}

The simplest way to extend the SM is to introduce the heavy Majorana neutrinos and allow for lepton number violation.
The introduced heavy Majorana neutrinos can simultaneously explain the tiny neutrino masses via seesaw mechanism and the baryon asymmetry of the Universe via leptogenesis.
In this paper, we explore the prospects for measuring CP violation in rare $W$ decays at the LHC, where the CP asymmetry between $W^- \to \ell_{\alpha}^- \ell_{\beta}^- (q \bar{q}^\prime)^+$ and $W^+ \to \ell_{\alpha}^+ \ell_{\beta}^+ (\bar{q}q^\prime)^{-}$ arises from the significant interference of contributions from two different Majorana neutrinos.
We find that, the CP asymmetry for fixed $\Delta\phi$ is basically independent of the Majorana neutrino mass in the mass range of our interest $10~\gev < \mn < 70~\gev$.
Taking $m_{N_{1}}=10~\gev$ and $\Delta\phi=\pi/2$ as an example, we investigate the possibility of measuring such CP violation at 14 TeV and 100 TeV LHC.
Although the signal cross section at 14 TeV LHC is too small to be detected experimentally, the high energy 100 TeV LHC will offer us a great opportunity to explore CP violation effects.
The measurement of such CP violation would provide us much important information about the underlying new physics.

\section*{Acknowledgements}

P.~C.~Lu, Z.~G.~Si, Z.~Wang and X.~H.~Yang thank the members of the Institute of theoretical physics of Shandong University for their helpful discussions.
This work is supported in part by National Natural Science Foundation of China (Grants No. 11875179, 11775130) and Natural Science Foundation of Shandong Province (Grant No. ZR2021QA040).

\begin{appendix}

\section{Formalism for rare $W$ decay }\label{appA}

The functions ${\cal T}_i$ ($i = 1, 2$) and ${\cal T}_{12}$ in Eq.~\ref{6} can be respectively expressed as
\begin{align}
\label{A1}
{\cal T}_i =& \left|D_{N_i}\left(p_N^2\right)\right|^2 \cdot {\cal F}
- {\rm Re} \left[D_{N_i}\left(p_N^2\right)D_{N_i}^\ast\left({p_N^\prime}^2\right)\right] \cdot {\cal I}\nn \\
&+ {\rm Im} \left[D_{N_i}\left(p_N^2\right)D_{N_i}^\ast\left({p_N^\prime}^2\right)\right] \cdot {\cal J} \; , \\
\label{A2}
{\cal T}_{12} =&  \left[D_{N_1}\left(p_N^2\right)D_{N_2}^\ast\left({p_N}^2\right) + D_{N_1}\left({p_N^\prime}^2\right)D_{N_2}^\ast\left({p_N^\prime}^2\right) \right] \cdot {\cal F}
   \nn \\
&- \left[D_{N_1}\left(p_N^2\right)D_{N_2}^\ast\left({p_N^\prime}^2\right) + D_{N_1}\left({p_N^\prime}^2\right)D_{N_2}^\ast\left(p_N^2\right)\right] \cdot {\cal I} \nn \\
&+ i \left[D_{N_1}\left(p_N^2\right)D_{N_2}^\ast\left({p_N^\prime}^2\right) - D_{N_1}\left({p_N^\prime}^2\right)D_{N_2}^\ast\left(p_N^2\right)\right] \cdot {\cal J} \; ,
\end{align}
where $p_N=p_2-p_1$ and $p_N^\prime=p_1-p_3$. $D_{N_i}\left(p^2\right)$ $(i=1,2)$ is the Breit-Wigner propagator and can be defined as
\begin{align}
\label{A3}
D_{N_i}\left(p^2\right) = \frac{1}{p^2-m_{N_i}^2+im_{N_i}\Gamma_{N_i}} \; ,
\end{align}
with $m_{N_i}$ and $\Gamma_{N_i}$ being the mass and total decay width of the two Majorana neutrinos $N_1$ and $N_2$.

The explicit expressions of ${\cal F}$, ${\cal I}$ and ${\cal J}$ introduced in Eq.~\ref{A1} and Eq.~\ref{A2} can be given by
\begin{align}
\label{}
{\cal F} = & 16 (p_4 \cdot p_3) \left[m_W^2 (p_5 \cdot p_2) + 2(p_5 \cdot p_1) (p_1 \cdot p_2) \right] \; , \\
{\cal I} = & 8 \biggl\{-(p_4 \cdot p_3) \left[m_W^2 (p_5 \cdot p_2)+2 (p_5\cdot p_1)(p_1\cdot p_2) \right] -(p_4\cdot p_2) \left[m_W^2 (p_5\cdot p_3)+2(p_5\cdot p_1)(p_1\cdot p_3) \right]   \nn \\
&+(p_2\cdot p_3)\left[m_W^2(p_4\cdot p_5)+2(p_4\cdot p_1)(p_5\cdot p_1) \right]\biggr\}   \; ,  \\
{\cal J} = & 8\left[m_W^2+2(p_5\cdot p_1) \right] \epsilon_{p_1 p_2 p_3 p_5} \; ,
\end{align}
where $\epsilon_{p_1 p_2 p_3 p_5}=\epsilon_{\mu \nu \rho \sigma}p_1^\mu p_2^\nu p_3^\rho p_5^\sigma$.

\end{appendix}

\end{document}